\newcommand{\cn}{{\mathrm{cn}} }
\newcommand{\dn}{{\mathrm{dn}} }
\newcommand{\sn}{{\mathrm{sn}} }

\newcommand{\tx}{\tilde{x}}
\newcommand{\tr}{\tilde{r}_\bot}

\documentclass{ws-mplb}

\begin{document}


%
\catchline{}{}{}{}{}
%

\title{Theory of Nonlinear Matter Waves in Optical Lattices}

\author{\footnotesize V.A. BRAZHNYI$^1$\footnote{E-mail: brazhnyi@cii.fc.ul.pt}\quad and V.V. KONOTOP$^{1,2}$\footnote{E-mail: konotop@cii.fc.ul.pt}}

\address{$^1$Centro de F\'{\i}sica Te\'{o}rica e Computacional, Universidade de Lisboa, 
Complexo Interdisciplinar, Av. Prof. Gama Pinto 2, Lisbon 1649-003, Portugal\\
$^{2}$
Departamento de F\'{\i}sica,
Universidade de Lisboa, Campo Grande, Ed. C8, Piso 6, Lisboa
1749-016, Portugal.}

\maketitle

\begin{history}
\received{(received date)}
\revised{(revised date)}
\end{history}

\begin{abstract}
We consider several effects of the matter wave dynamics which can be observed in  Bose-Einstein condensates embedded into optical lattices. For low-density condensates we derive approximate evolution equations, the form of which depends on relation among the main spatial scales of the system. Reduction of the Gross-Pitaevskii equation to a lattice model (the tight-binding approximation) is also presented.  Within the framework of the obtained models we consider modulational instability of the condensate, solitary and periodic matter waves, paying special attention to different limits of the solutions, i.e. to smooth movable gap solitons and  to strongly localized discrete modes. We also discuss how the Feshbach resonance, a linear force,  and lattice defects affect the nonlinear matter waves.   
\end{abstract}

\section{Introduction}

The first experimental realization of the Bose-Einstein condensate\cite{Anderson} (BEC) resulted in explosion of theoretical and experimental research devoted to the new state of matter, which now attracts a great deal of attention of the physical community. 
The studies have become really interdisciplinary, involving  scientists working not only in atomic physics, but also in nonlinear optics, condensed matter physics, dynamical systems, differential equations, etc. 
Analogies between the theory of nonlinear matter waves in BECs and earlier developments in other areas of the nonlinear science significantly accelerate the research progress. 
The major part of the  knowledge available today is gathered  in a number of recent reviews\cite{reviews}.

It is widely accepted that generation, dynamics and management of nonlinear matter waves are the most interesting and important issues of the mean-field theory of a BEC. 
In particular, three types of relatively stable coherent structures play the main role in the theory. They are {\em bright solitons},  characterized by atomic density distribution localized in space; {\em dark solitons}, representing holes in the otherwise homogeneous atomic distributions; and nonlinear {\em periodic waves}. 
Although these excitations can be observed experimentally in  BECs trapped by a magnetic field\cite{Khay,bright,dark}, due to the experiment\cite{Ander} it became clear that the matter wave dynamics can be drastically changed and appreciably enriched if an optical lattice (OL) is imposed on the condensate (in addition to or instead of a harmonic trap). 
That is why BECs loaded into OLs, which are created by laser beams, attract especial attention.
A large diversity of remarkable effects has already been observed experimentally. 
Among them we mention Bloch oscillations of BECs\cite{Ander,Morsch}, instability of nonlinear matter waves\cite{instab}, 
Landau-Zener tunneling\cite{LZ}, superfluid to Mott-insulator phase transition\cite{Greiner}, compression of a condensate\cite{lens}, and gap matter solitons\cite{gap_sol}. 

Behavior of a BEC strongly depends on parameters of an OL which in its turn can be controlled by intensities and/or by geometry of laser beams.  
In other words, the matter wave dynamics can be effectively manipulated by external fields, the problem having primary importance for a number of BEC applications, like atomic interferometry\cite{Wright}, quantum phase transitions\cite{Ueda}, matter wave amplification\cite{amplific}, atom laser\cite{laser}  and others.
 
The scope of this brief review does not allow us to cover all important issues. 
Instead, we deal with a few topics of the mean-field theory of BECs embedded in OLs, only mentioning, in Conclusion, other important issues.
\section{Mean-filed approximation}
\label{lattice}

Quantum description of a BEC in the approximation of two-body interactions is provided by the nonlinear Schr\"{o}dinger (NLS) equation for the field operators  ${\bf \Psi}\equiv {\bf \Psi}({\bf r}, t)$\cite{LL}:
\begin{eqnarray}
\label{eq:QGP}
	i\hbar\frac{\partial {\bf \Psi}({\bf r},t)}{\partial t}=\left[-\frac{\hbar^2}
{2{\cal M}}\Delta+V_{ext}({\bf r})+\int d^3{\bf r}^{\prime} {\bf \Psi}^{\dag}({\bf r}^{\prime},t) V({\bf r}^{\prime}-{\bf r}){\bf \Psi}({\bf r}^{\prime},t)\right]
{\bf \Psi}({\bf r},t)\, ,
\end{eqnarray}
where $V_{ext}({\bf r})$ is an external trap potential,  ${\cal M}$ is an atomic mass, and $ V({\bf r})$ describes interaction between two atoms. 
In the present paper the consideration will be restricted to  local interactions, when one can approximate $V({\bf r})= g_0\delta({\bf r})$ with $g_0=4\pi \hbar^2a_s/{\cal M}$.
Moreover, at zero temperature, the operator ${\bf \Psi}$ can be represented in the form ${\bf \Psi}=\Psi+\hat{\psi}$, where $\Psi$ is an algebraic function, called an order parameter (or macroscopic wave function) and the operator $\hat{\psi}$ describes quantum fluctuations\cite{gardiner}.
Then, neglecting quantum fluctuations\cite{gardiner}, the dynamics of the condensate can be described by the macroscopic wave function $\Psi$ which solves the Gross-Pitaevskii (GP) equation\cite{GP}  
\begin{equation}
\label{GPE}
i\hbar \frac{\partial\Psi}{\partial t}=-\frac{\hbar^2}
{2{\cal M}}\Delta\Psi+V_{ext}({\bf r})\Psi+g_0|\Psi|^2\Psi\, 
\end{equation}
and is normalized to the number of condensed atoms ${\cal N}$:
\begin{eqnarray}
	\label{norm}
	\int |\Psi|^2d^3{\bf r}={\cal N}.
\end{eqnarray}

In the case of a BEC embedded in an OL the external potential usually consists of two parts: $V_{ext}({\bf r})=V_{T}({\bf r})+V_{L}({\bf r})$.
The first term is a harmonic trap given by $V_{T}({\bf r})=\frac {\cal M}2 \left(\omega_{\bot}^2{\bf r}_\bot^2+\omega_0^2x^2\right)$ where  ${\bf r}=(x,{\bf r}_\bot)$, ${\bf r}_\bot=(y,z)$, and $\omega_{\bot}$ and $\omega_0$ are a transverse and axial harmonic oscillator frequencies (for the sake of simplicity the consideration will be restricted to radially symmetric traps). 
Respectively, one can introduce two spatial scales $a_{\bot}=\sqrt{\hbar/m\omega_{\bot}}$ and $a_0=\sqrt{\hbar/m\omega_0}$, characterizing transverse and longitudinal extensions of the atomic cloud.
The second term is a periodic potential which, almost everywhere through this paper is considered  to be one-dimensional (1D), $V_L({\bf r})\equiv V_L(x)=V_L(x+\Lambda)$, with a period $\Lambda$.

\begin{figure}[ht]
\centerline{\psfig{file=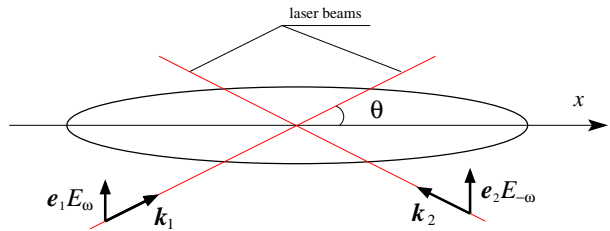,width=8cm}}
\vspace*{8pt}
\caption{Typical geometry of a BEC in an OL. Here  ${\bf k}_1$, ${\bf k}_2$, and the $x$-axis lie in the same plane.  $|{\bf k}_1|=|{\bf k}_2|=k$, $|{\bf e}_1|=|{\bf e}_2|=1$ and ${\bf e}_1\parallel {\bf e}_2$.}
\label{lattice-fig-one}        
\end{figure}

Interaction of a single atom, having an electric dipole moment ${\bf d}$, with an electric field ${\bf E}$ is  described by the Hamiltonian $H_{d}=-{\bf d\cdot E}$, which results in the shift of the energy level (see e.g.\cite{PetSmith}): $V_{L}({\bf r})=-\frac{1}{2}\alpha^\prime\langle {\bf E}^2\rangle_t$ where $\langle\cdot\rangle_t$ stands for time averaging and $\alpha^\prime$ is the real part of the polarizability. 
Let us consider two monochromatic laser beams, 
${\bf E}={\bf e}_1E_{\omega}e^{i\omega_L t}+{\bf e}_2E_{-\omega}e^{-i\omega_L t}$, where $\omega_L$ is a frequency, ${\bf e}_j$ are polarization verctors, and $E_{\omega}=\bar{E}_{-\omega}$ (an overbar  stands for the complex conjugation),
in a geometry\cite{Grynb} typical for many experimental settings\cite{LZ} and shown in Fig.~\ref{lattice-fig-one},  
Then, neglecting an insignificant  constant, one obtains the  lattice potential  
\begin{eqnarray}
\label{eq:linV}
V_L(x)=V_0\sin^2(\kappa x)
\end{eqnarray}
where $\kappa=k\cos\theta=\pi/\Lambda$ with $k=2\pi/\lambda$, $\lambda$ is the laser wavelength,  $V_0=\hbar\Omega_R^2/(\omega_L-\Omega_0)$ is the amplitude of the OL, $\Omega_R=|{\bf d}\cdot{\bf E}|/\hbar$ is the Rabi frequency, and $\hbar\Omega_0$ is the energy difference between the ground and first excited atomic states. Hereafter we neglect losses assuming that the laser beams are far-off-resonant.

The BEC dynamics is characterized by one more relevant length scale -- the healing length $\xi=(8\pi n|a_s|)^{-1/2}$,  where $a_s$ the s-wave scattering length and $n\propto  {\cal N}/a_{\bot}^2a_0$ is a mean particle density. 
Thus we collect four\footnote{$|a_s|$ being of order of a few nanometers is much smaller than other characteristic scales.} characteristic scales  which will define the dynamics of a BEC in an OL: $\{a_0,a_{\bot},\Lambda,\xi\}$\footnote{Although,  we introduced $\Lambda$ through the laser wavelength, bellow we use this notation in a generalized sense of a period of the potential. For example, in the case (\ref{eq:pot}) $\Lambda=2{\rm K}(m)/\kappa$.}. 
Alternatively, one can consider a hierarchy of the respective energy scales: $\{\hbar\omega_\bot,\hbar\omega_0,E_R,g_0 n\}$, where $E_R=\hbar^2\kappa^2/(2{\cal M})$ is the recoil energy.

It worth pointing out that by means of using several laser beams one can create more sophisticated periodic potentials. From the theoretical point of view  potentials expressed through  Jacobi elliptic functions play a special role, because they may allow obtaining exact solutions\cite{24} (see also Section~\ref{periodic}). In particular, we will be interested in the potential
\begin{eqnarray}
\label{eq:pot}
V_L(x)=V_{0}\sn^2(\kappa x\mid m)
\end{eqnarray}
which is reduced to (\ref{eq:linV}) in the limit $m\to 0$. Hereafter we use the standard\cite{AbSteg} notations
pq$(\kappa x\mid m)$, where p=s, c, or d, and ${\rm q=n}$, for the Jacobi elliptic functions with the parameter $0\leq m\leq 1$. Indeed, using the expansion of the elliptic sine\cite{AbSteg} in terms of the small parameter ${\rm q}=\exp\left(-\pi\frac{{\rm K}(1-m)}{{\rm K}(m)}\right)\ll 1$, where ${\rm K}(m)$ (and ${\rm E}(m)$ below) is a complete elliptic integral of the first (respectively second) kind, one can approximate
\begin{eqnarray*}
V_L(x)\approx\frac{2 V_0 \pi^2{\rm q}}{m{\rm K}^2(m)(1-{\rm q})^2}
\left[1-(1-2{\rm q})\cos\frac{\pi\kappa x}{{\rm K}(m)}- 2{\rm q} 
\cos\frac{2\pi\kappa x}{{\rm K}(m)}\right].
\end{eqnarray*}
In practice this means that by using four laser beams one can approximate potential (\ref{eq:pot}) for a rather large interval of $m$, ranging from zero up to approximately $0.9$ with accuracy higher than 99\%.

\section{One-dimensional models}
\label{meanfield}

Let us consider a cigar-shaped condensate: $a_\bot\ll a_0$, at a relatively low density, when the energy of two body interactions is much less than the kinetic energy in the transverse direction, i.e.  when
\begin{eqnarray}
	\label{small_pa}
\epsilon^2=\frac{a^2_\bot}{\xi^2}\sim \frac{{\cal N} a_s}{a_0}\ll 1.
\end{eqnarray}
Then, the system becomes effectively 1D\cite{Humberto}.
In literature there exist several approximations reducing the original GP equation to an effective 1D GP (or NLS) model. 
A self-consistent way, allowing one to consider different relations among parameters and to control magnitudes of neglected terms, consists in use of the multiple-scale expansion with respect to the small parameter $\epsilon$ given by (\ref{small_pa}). 
We outline the method for a specific relation among the parameters, referring for more details to\cite{KS1,VVKBook,BaizKonSal02}, and list other relevant cases. 

To this end we scale out the variables
$\tilde{t}=\omega_{\bot} t/2$, $\tilde{x}=x/a_{\bot}$, $\tilde{\bf r}_\bot={\bf r}_\bot/a_{\bot}$, and the
wave function $\tilde{\Psi}=a_{\bot}|a_s|^{1/2}\Psi$, and  consider a long trap with a rapidly varying OL, what is  quantified by the relations $a_{\bot}\sim \Lambda \sim\epsilon\xi\lesssim \epsilon^2a_0$. 
In this case GP equation (\ref{GPE}) is rewritten as  
\begin{eqnarray}
\label{GPE1}
\begin{array}{l}
\displaystyle{
i\partial_{\tilde{t}}\tilde{\Psi} =\left({\cal L}_\bot+{\cal L}_0 +8\pi\sigma|\tilde{\Psi}|^2\right)\tilde{\Psi},
}
\\
\displaystyle{
{\cal L}_\bot=-\Delta_\bot + \tilde{r}^2_\bot,\quad {\cal L}_0=-\partial_{\tilde{x}}^2 +\nu^2 \tilde{x}^2+\gamma^2U(\gamma \tilde{x}).
}
\end{array}
\end{eqnarray}
where $\sigma=$sign$(a_s)$, $\nu=\omega_0/\omega_\bot$  is the aspect ratio of the trap,   $U(\gamma\tilde{x})\equiv V_{L}(x)/E_R$ is a dimensionless periodic potential varying on a unit scale, $U(\gamma\tilde{x})=U(\gamma(\tilde{x}+\tilde{\Lambda}))$, $\tilde{\Lambda}=\Lambda / a_\bot\sim 1$, and 
$ \gamma^2=\frac{2E_R}{\hbar\omega_\bot}=\frac{4\pi^2}{\tilde{\Lambda}^2}$. 

Next we consider the linear eigenvalue  problems 
\begin{eqnarray}
	\label{lin_prob}
 {\cal L}_0\, \varphi_{\alpha, q}(\tilde{x})={\cal E}_{\alpha, q}\, \varphi_{\alpha, q}(\tx),
 \qquad
{\cal L}_\bot\, \phi_{lm}(\tilde{{\bf r}}_\bot)=E_{lm}\, \phi_{lm} (\tilde{{\bf r}}_\bot), 
\end{eqnarray}
where the indexes $\alpha $ and $q$ stand for a number of a band and for a wave vector inside the first Brillouin zone (BZ), and $l$ and $m$ stand for radial and magnetic quantum numbers. 
The operator ${\cal L}_0$ has a discrete spectrum, which however approximates the band spectrum in the limit $\nu\ll \epsilon$, what justifies the use of the ``band'' terminology.  
Now $\varphi_{\alpha ,q}(x)$ is approximated by a Bloch function (BF) of the respective periodic potential, i.e. by a solution of the eigenvalue problem  
\begin{eqnarray}
\label{Bloch}
	-\frac{d^2\varphi_{\alpha, q}(x)}{dx^2}+U(x)\varphi_{\alpha, q}(x)={\cal E}_{\alpha, q}\varphi_{\alpha, q}(x),
\end{eqnarray}
where taking into account that $a_{\bot}\sim \Lambda$ 
we put $\gamma=1$ and provisionally dropped tildes.
Below we will restrict the consideration to the states bordering an edge of the first BZ where 
$q=q_0= 2\pi/\tilde{\Lambda}$, and respectively will use  abbreviated notations $\varphi_\alpha (x)\equiv \varphi_{\alpha, q_0}(x)$ and ${\cal E}^{(\alpha)}= {\cal E}_{\alpha, q_0}$. Also we will be interested only in evolution of the background transverse state ($l=m=0$) corresponding to $\phi_{00}(\tilde{{\bf r}}_\bot)=\exp(-\tr^2/2)/\sqrt{\pi}$ and  to $ E_{00}=2$.\footnote{Strictly speaking inter-level transitions are not forbidden. They however can be originated by lattice perturbations or by initial conditions which will not be considered here.} All eigenfunctions are considered to be normalized to one.

Now we look for a solution of Eq.(\ref{GPE1}) in a form 
$\tilde{\Psi}=\epsilon\psi_1+\epsilon^2\psi_2+\cdots,
$ where $\psi_j$ ($j=1,2,\ldots$) are functions of $\tilde{{\bf r}}_\bot$ and of scaled variables $t_p=\epsilon^p \tilde{t}$ and $x_p=\epsilon^p \tilde{x}$, regarded as independent. It follows from (\ref{norm}) that $\psi_1$ is normalized as follows $\int|\psi_1|^2d^3\tilde{{\bf r}}=\frac{a_0}{a_\bot}$ what means that $|\psi_1|^2=O(1)$. Then the leading order of the solution can be searched in a form of the modulated ground state 
\begin{eqnarray}
\label{gr_st}
\psi_1=\frac{1}{\sqrt{\pi}} {\cal A}(x_1,t_2) e^{-i({\cal E}^{(\alpha)}+2)t_0}e^{-\tr^2/2}\varphi_{\alpha}(x_0).	
\end{eqnarray}
Here ${\cal A}(x_1,t_2)$ (as well as ${\cal B}_n(x_1,t_2)$ below) is a slowly varying envelope amplitude, where by convention we indicate only the most rapid variables (i.e., for example, ${\cal A}(x_1,t_2)$ means that ${\cal A}$ depends on all $\{x_1,x_2,\ldots\}$ and $\{t_2,t_3,\ldots\}$).
  
After substituting the expansions into (\ref{GPE1}) one collects the terms at each  order of $\epsilon$ (for the set of so obtained equations see e.g.\cite{KS1,BaizKonSal02}). 
The first order  equation is satisfied by ansatz (\ref{gr_st}).
In the second order one obtains
\begin{eqnarray}
	\label{psi2}
	\psi_2=\frac{1}{\sqrt{\pi}}e^{-\tr^2/2}e^{-i({\cal E}^{({\alpha})}+2)t_0} \sum_{\alpha^\prime\neq \alpha} {\cal B}_{\alpha^\prime \alpha} (x_1,t_1) \varphi_{\alpha^\prime} (x_0)
\end{eqnarray}
where 
$
{\cal B}_{\alpha_1\alpha}=\frac{\Gamma_{\alpha_1\alpha}}{{\cal E}^{(\alpha_1)}-{\cal E}^{(\alpha)}} \partial_{x_1} {\cal A}
$
and
$
\Gamma_{\alpha_1\alpha}=-\int_0^{\tilde{\Lambda}}\bar{\varphi}_{\alpha_1}\frac{d\varphi_{\alpha}}{dx_0} dx_0.
$
Thus $\psi_2$ always exists when the ground state  is excited (see Fig.~\ref{modinstab}). 
That is why, following the nonlinear optics terminology\cite{optics}, we call $\psi_2$ a {\em companion mode}. 
Also it follows from the second order equation that
$\partial_{t_1} {\cal A}=0$, i.e. ${\cal A}\equiv {\cal A}(x_1,t_2)$. 

Eliminating secular terms from the third order equation we obtain
\begin{eqnarray}
\label{nls1D_A}
i\partial_{t_2} {\cal A} +(2M_\alpha)^{-1}\partial_{x_1}^2 {\cal A}- \chi_\alpha\sigma|{\cal A}|^2{\cal A} =0,	
\end{eqnarray}
where  the {\em effective nonlinearity}, $\chi_\alpha$, and the {\em inverse effective mass}, $M_\alpha$, are given by 
\begin{eqnarray}
\label{eff_mass}
\chi_\alpha=2\int_0^{\tilde{\Lambda}} |\varphi_{\alpha}(x_0)|^4dx_0, \quad\mbox{and}\quad\frac{1}{M_\alpha}=2+2\sum_{\alpha_1\neq \alpha}\frac{|\Gamma_{\alpha\alpha_1}|^2}{{\cal E}^{(\alpha_1)}-{\cal E}^{(\alpha)}}.
\end{eqnarray}
By substitutions ${\cal A}\to \sqrt{2/\chi_\alpha}\psi$, $t_2\to t$, $x_1\to x$, and $M_\alpha\to M$ Eq. (\ref{nls1D_A}) is reduced to a dimensionless NLS equation
\begin{eqnarray}
\label{nls1D}
i \partial_t \psi =-\frac{1}{2M}\partial^2_x \psi+ 2\sigma|\psi|^2\psi\,.	
\end{eqnarray}
To conclude we have to clarify the meaning of the coefficient $M_\alpha^{-1}$. 
This can be done\cite{optics}, by means of the so-called ${\bf kp}$-method,  known in the solid state physics\cite{solid}.  Namely, it turns out that in the limit $a_0/\Lambda\to \infty$ one has\cite{KS1} $M_\alpha^{-1}=d^2{\cal E}_{\alpha, q}/dq^2$. 

By analogy one can consider other relevant cases as follows (notice that in each case the operator ${\cal L}_0$ is redefined):
  
-- {\em Short trap and rapidly varying optical lattice}:
$a_\bot\sim \Lambda \sim \epsilon\xi\sim\epsilon a_0$. 
Wave function satisfies  the equation (all tildes are suppressed)
\begin{eqnarray}
\label{nls1dcase2}
i\partial_t\psi=-\frac{1}{2M}\partial^2_{x}\psi + \nu^2x^2\psi +2\sigma |\psi|^2\psi 	
\end{eqnarray}
The eigenfunctions $\varphi_\alpha$ are now the exact BFs solving Eq. (\ref{Bloch}).

--  {\em Long trap and smooth optical lattice}: $a_\bot\sim \epsilon \Lambda \sim\epsilon\xi\lesssim\epsilon^2a_0 $. Wave function is governed by  the equation (in the final equation all tildes are suppressed)
\begin{eqnarray}
\label{case3nls} 
i\partial_t\psi=-\partial^2_x\psi+ U(x)\psi +2\sigma |\psi|^2\psi.
\end{eqnarray}
Now the periodic potential has the scale equal to (or larger than) the healing length, and thus  excitations in a BEC have also scales of order of the lattice period. It is also assumed that $\gamma=\epsilon\tilde{\gamma}$, where $\gamma$ is the aspect ratio and $\tilde{\gamma}=(2a_\bot/\Lambda)(a_0/{\cal N}|a_s|)^{1/2}=O(1)$, and that $\gamma^2U(\gamma \tx)\equiv \epsilon^2\tilde{U}(x_1)$.  
   
-- {\em Short magnetic trap and smooth optical lattice}: 
$a_\bot\sim\epsilon \Lambda \sim\epsilon\xi\sim\epsilon\, a_0$. Wave function is described by the equation
\begin{equation}
\label{nls1_case4} 
i\partial_t\psi=-\partial^2_x\psi+ \nu^2x^2\psi+U(x)\psi+2\sigma|\psi|^2\psi.
\end{equation}
 
Below we use one of these models depending on the effect we are interested in. 
\section{Modulational instability}
\label{instability}

As it is clear from the preceding section, the eigenfunctions of the linear problem  (\ref{Bloch}) play an important role in the theory of low density BECs. 
When inter-atomic interactions are neglected, they represent renormalized wave functions of single atoms, whose evolution in time is trivial. 
Interactions drastically change this situation, leading in some cases to instability of the ground state. 
This is a phenomenon usually referred to as {\em modulational instability} (MI) and 
is well studied in hydrodynamics, plasma physics, nonlinear optics, etc.\cite{rew}. 
 
\begin{figure}[h]
\centerline{\epsfig{file=ek.eps,width=6.5cm}\quad
\epsfig{file=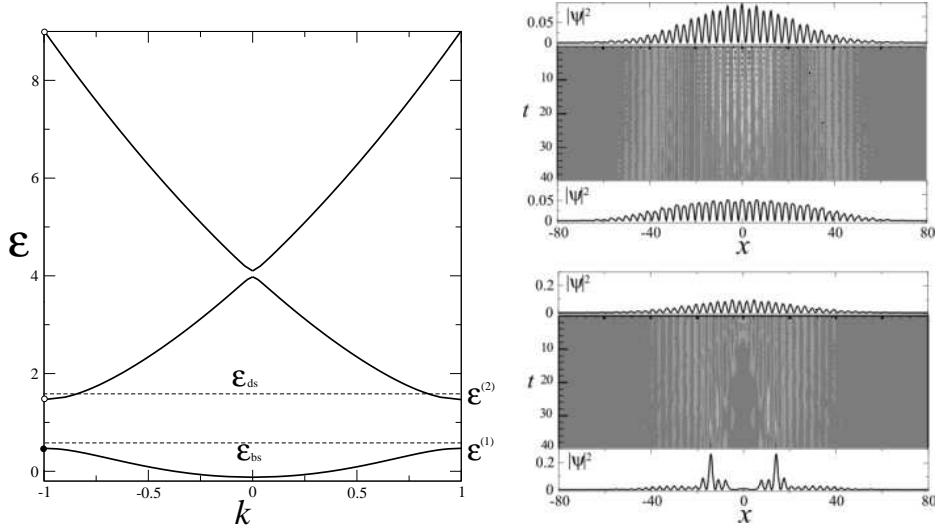,width=5.5cm}}
\vspace*{8pt}
\caption{Left panel: The band gap structure of problem (\ref{Bloch}) with $A=1$. In a BEC with a positive scattering length  ${\cal E}^{(1)}$ (respectively $M_1<0$) and ${\cal E}^{(2)}$ (respectively $M_2>0$) correspond to unstable and stable Bloch states. If the main mode, $\psi_1$ (see (\ref{gr_st})), is excited against a carrier wave background $\varphi_1(x)$ (the filled in symbol) then the companion modes $\psi_2$ (see (\ref{psi2})) are excited with the same wavevector but belonging to different zones (open symbols). The bright soliton frequency ${\cal E}_{{\rm bs}}$ (the lowest dashed line) is shifted toward the gap. In the small amplitude limit the shift is given by $\eta^2/(2|M|)$ (see (\ref{bright})). The upper dashed line shows the energy of the dark soliton, shifted toward the allowed band (in the small amplitude limit the shift is given by $\eta^2/|M|$ (see  (\ref{dark})).
Right panels: Numerical solution of 3D GP equation (\ref{GPE1}) with cigar shaped geometry and 1D OL in the $x$-direction as in the left panel. 
As an initial condition we use either $\varphi_1(x)$, well approximated by  $\sqrt{2}\sin(x)$ (the right bottom panel), or $\varphi_2(x)$, approximated by  $\sqrt{2}\cos(x)$ (the top panel), both modulated by the Gaussian function $e^{-r^2/2-\nu x^2/2}$.}
\label{modinstab}        
\end{figure}

In order to develop a theory of MI of a BEC embedded in an OL we, following Refs.\cite{KS1,BaizKonSal02}, consider first the model (\ref{nls1D}) and look for a solution in a form of the plane wave  $\psi=(\rho+ae^{i(\Omega t-{\cal K}x)}+be^{-i(\Omega t-{\cal K}x)})e^{-i\rho^2 t}$, where $|a|$, $|b|\ll \rho$. Linearizing (\ref{nls1D}) with respect to $a$ and $b$ one obtains the dispersion relation $\Omega^2=Z(Z+4\sigma\rho^2)$, where $Z={\cal K}^2/2M$,  from which it follows that $\Omega$ has an imaginary component, and thus the plane wave is unstable, if  
\begin{equation}
\label{mi_1}
Z(Z+4\sigma \rho^2)<0.
\end{equation} 
This condition is satisfied for large wavelengths (small wavevectors, $|{\cal K}|<2\rho\sqrt{|M|}$) of the perturbation if $\sigma M<0$. Taking into account that the effective mass has different signs at different edges of the forbidden gap (see the example in Fig.\ref{modinstab}) we conclude that one of the two matter waves bordering the gap is modulationally unstable.

To explain MI also in a different way  we consider long wavelength excitations and substitute $\psi=\rho\exp(-i\Omega t)$ in (\ref{nls1D}). This gives a dispersion relation $\Omega=\sigma\rho^2$. Recalling that  $\psi$ is an envelope, the total frequency of the wave is given by ${\cal E}^{(\alpha)}+\epsilon^2\Omega$ (see (\ref{gr_st}) where 2 must be omitted in the 1D case). 
Then, if $\sigma>0$, for the upper gap edge ${\cal E}^{(2)}$  corresponding to the positive effective mass, $M>0$,
the energy shift due to the nonlinearity is positive, i.e. the total frequency falls into the allowed band (${\cal E}_{{\rm ds}}$ Fig.~\ref{modinstab}). 
At the edge ${\cal E}^{(1)}$, where $M<0$, the energy is shifted upward  in the gap (${\cal E}_{{\rm bs}}$ Fig.~\ref{modinstab}), where plane wave cannot exist, what means instability.
If $\sigma<0$,  the same arguments are applied to the ${\cal E}^{(1)}$ and ${\cal E}^{(2)}$ exchanged.

For smaller wavelengths of perturbation MI can be studied numerically\cite{KS1}. 
In the right panels in Fig.\ref{modinstab} we show the results of such study of the scaled out 3D GP equation  with the potential  
\begin{eqnarray}
\label{pot_cos}
U(x)=A\cos(2x).
\end{eqnarray}
The upper and lower panels show respectively stable and unstable evolution of the Bloch states bordering the edges of the first lowest gap. The localized pulses emerging  from the periodic wave are the bright {\em matter solitons} (see the next section).

The fact that an OL can result in the instability of an otherwise stable BEC has recently been confirmed experimentally\cite{instab}. 
  
The above analysis of the MI can be expanded   to 2D and 3D periodic potentials\cite{BaizKonSal02}.
In higher dimensional cases the inverse effective mass $M_{\alpha}^{-1}$ is replaced by the tensor of inverse masses $M^{-1}_{\alpha,\beta\gamma}\equiv\partial_\beta \partial_\gamma {\cal E}_{\alpha, {\bf k}}$ where $\partial_\beta=\frac{\partial}{\partial r_\beta}$, ${\bf r}=(r_1,r_2,r_3)=(x,y,z)$, and $\beta$, $\gamma=1,2,3$. When the multidimensional lattice potential is separable, say when $V_L({\bf r})=\sum_{j=1}^{3} V_j\cos(\kappa_j r_j)$, the tensor $M^{-1}_{\alpha,\beta\gamma}$ is diagonal, and the analog of model (\ref{nls1D}) reads\cite{BaizKonSal02}
\begin{equation}
\label{3dMI}
i\partial_t \psi =-\frac 12 \sum_j 	M^{-1}_{\alpha,jj} \partial_j^2 \psi + 2\sigma |\psi|^2\psi.
\end{equation}

\begin{figure}[h]
\centerline{\epsfig{file=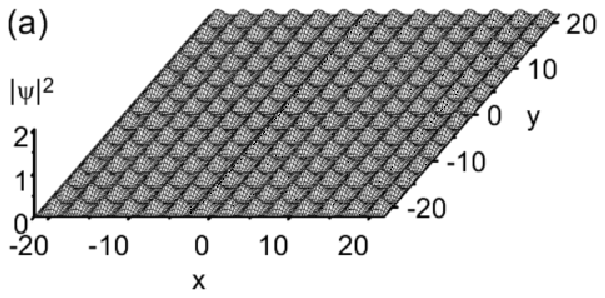,width=6cm}\quad \epsfig{file=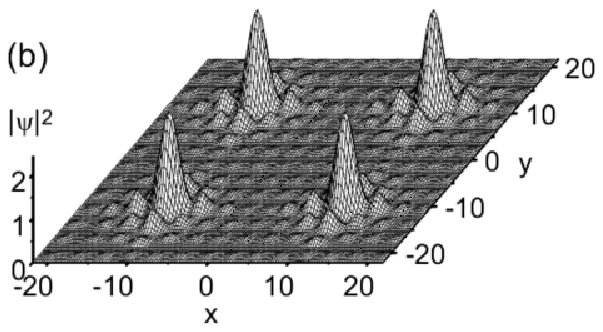,width=6cm}}
\vspace*{8pt}
\caption{The example (from Ref.$^{27}$) of unstable evolution of BEC atomic distribution in a 2D OL
according to equation (\ref{3dMI}) with $V_1=V_2=1.0, \ \kappa_1=\kappa_2=2.0, \ \sigma=1/2, \ L=14\pi. \ $
(a) The initial state:  $\psi(x,y,0) = 0.5 \sin(x) \sin(y)$ perturbed by
$\delta \psi (x,y) = 0.01 \sin(0.125 x) \sin(0.125 y)$.
(b) Formation of soliton-like excitations due to MI at $t=85$.
}
\label{modinst2}        
\end{figure}

Now the dynamics is more rich since BFs corresponding to different spatial dimensions contribute to the wave dynamics. The parameter $Z$, introduced in (\ref{mi_1}) has now the form $Z=\sum_j M_{\alpha, jj}^{-1}{\cal K}_j^2$. Depending on the curvature of the energy surface ${\cal E}_{\alpha, {\bf k}}$ in the given spatial direction, $\beta$, the respective band edge can be called either stable, if $\frac{\partial^2 {\cal E}_{\alpha, {\bf k}}}{\partial k_\beta^2}\sigma >0$, or unstable, if $\frac{\partial^2 {\cal E}_{\alpha, {\bf k}}}{\partial k_\beta^2}\sigma <0$. In Fig.\ref{modinst2}  we show unstable evolution of the BF in the two spatial dimensions (results from Ref.\cite{BaizKonSal02}). The figure clearly demonstrates emergence of a coherent structure consisting of spatially localized matter waves. The symmetry of the emerging structure is defined by the symmetry of the lattice and by the characteristic wavevector of the most unstable mode (see\cite{BaizKonSal02} for the details).  Existence of such structures leads to a supposition about possibility to generate 2D and 3D matter gap solitons. Such solitons have indeed been observed in numerical studies\cite{OstKiv}.
\section{Matter solitons and intrinsic localized modes}
\label{soliton}

MI instability considered above is not the only important consequence of the existence of the effective mass in the theory. Depending on the sign of $M\sigma$ an OL acts as a focusing medium, if $M\sigma<0$, or defocusing medium, if $M\sigma>0$.  
Taking into account that by using a moving lattice one can obtain either of two signs of the effective mass in the case when the nonlinearity dominates dispersion, the lattice can be viewed as a compressor of the matter waves\cite{MM}, which was recently achieved  experimentally\cite{lens}. 

When nonlinearity and dispersion are in balance, creation of stable structures, termed solitons is possible. 
In particular,  NLS equation (\ref{nls1D}) possesses\cite{IST} either a {\em bright soliton} solution
\begin{equation}
	\label{bright}
	\psi_{{\rm bs}}(x,t)= \frac{\eta}{\sqrt{2|M|}}\exp\left(i \frac{\eta^2}{2M}t\right)\cosh^{-1}(\eta x)
\end{equation}
if $M\sigma<0$, or a {\em dark soliton} solution
\begin{equation}
	\label{dark}
	\psi_{{\rm ds}}(x,t)=\frac{\eta}{\sqrt{2|M|}}\exp\left(-i \frac{\eta^2}{M}t\right)\tanh(\eta x)
\end{equation}
if $M\sigma>0$. The parameter $\eta$ characterizes widths and amplitudes of the solitons.  
It is natural to refer to these solutions as {\em matter gap solitons}. 
Well known in nonlinear optics of photonic crystals\cite{optics} matter gap solitons 
have been predicted in Ref.\cite{SZ}. 
Very recently bright solitons in a condensed cloud of $^{87}$Rb atoms have been generated experimentally\cite{gap_sol}.

As one can see from Eqs. (\ref{nls1D}) and (\ref{nls1dcase2})   one of the main features introduced by an OL with a small lattice constant is a change of the spectrum.
This is expressed in terms of the effective mass (\ref{eff_mass}) which can change its sign.
Thus even in the case of repulsive (attractive) atomic interactions in the presence of OL one can excite bright (dark) matter solitons, what is impossible in  homogeneous condensates. 
As it follows from the multiple-scale analysis  solitons (\ref{bright}), (\ref{dark}) are characterized by the following main features:
(i) They are low-amplitude, having frequencies close to the gap edge (shifted toward  forbidden and allowed bands, respectively); (ii) They extend over many lattice periods;   (iii) They cannot be excited without companion modes, what  contributes to instability and finite life-time of these entities;
(iv) They are not exact solutions of the GP equation, and as such can be deformed during propagation what results in their decay with time (for radiative losses of matter gap solitons see Ref.\cite{YSR}).

\subsection{Tight-binding approximation}
\label{TBA}

If a gap, i.e. $|{\cal E}^{(2)}-{\cal E}^{(1)}|$, is large enough, what happens if the  potential relief is deep enough, then when the energy shift $|{\cal E}-{\cal E}^{(\alpha)}|$ outwards the gap edge ${\cal E}^{(\alpha)}$ increases, a matter soliton does not disappear but becomes strongly localized on a very few lattice cites\cite{TrombSmerz,ABDKS}. 
Now one cannot use model (\ref{nls1D}) any more, and has to explore Eq. (\ref{case3nls}) describing the situation where the healing length is of order of the lattice period. 
The respective approach is often referred to as a {\em tight-binding approximation}\cite{TrombSmerz,ABDKS}. We introduce it following Ref.\cite{AKKS} where a mathematical background for the method was elaborated. 

\begin{figure}[h]
\centerline{\epsfig{file=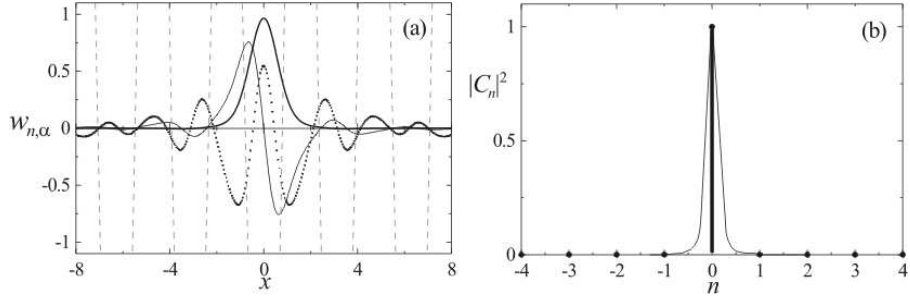,width=12cm}}
\caption{(a) WFs of the three lowest bands for the potential (\ref{pot_cos}) with $A=5$. The WF of the first zone (bold solid line; thin solid line and dotted line show WFs of the second and the third zones, respectively)  coincides in the leading order with the shape of a Wannier soliton. (b) Atomic distribution over the sites in a discrete soliton corresponding to the Wannier soliton.  
} 
\label{fig:WF}
\end{figure}

Periodicity of the lattice potential implies  periodicity of the eigenvalues of (\ref{Bloch}): ${\cal E}_{\alpha,q+q_0}={\cal E}_{\alpha,q}={\cal E}_{\alpha,-q}$ where $q_0=2\pi/\tilde{\Lambda}$ what allows one to expand ${\cal E}_{\alpha,q}$ in the Fourier series:
${\cal E}_{\alpha,q}= \sum_{n}\hat{\omega}_{n,\alpha}e^{i qn\tilde{\Lambda}}$,  
where $\hat{\omega}_{n,\alpha } =\frac{\tilde{\Lambda}}{2\pi}\int_{-q_0/2}^{q_0/2}{\cal E}_{\alpha,q}e^{-i qn\tilde{\Lambda}}dq$.
Next, we introduce the Wannier functions\cite{kohn} (WFs)
\begin{eqnarray}
\label{wannier1}
w_{n,\alpha}(x)
=\sqrt{\frac{\tilde{\Lambda}}{2\pi}}
\int_{-q_0/2}^{q_0/2}\varphi_{\alpha, q}(x) e^{-i qn\tilde{\Lambda}}dq,
\quad
\varphi_{\alpha,q}(x)=\sqrt{\frac{\tilde{\Lambda}}{2\pi}}\sum_n w_{n,\alpha}(x)e^{iqn\tilde{\Lambda}}
\end{eqnarray}
(here $n$ is an integer). Some examples of WFs are shown in Fig.~\ref{fig:WF}a. Taking into account that WFs are normalized, $\int_{-\infty}^{\infty}\bar{w}_{n,\alpha}(x)w_{n',\alpha'}(x)\,dx=\delta_{n,n'}\delta_{\alpha,\alpha'}$,
make up a complete set, $\sum_{n\alpha}\bar{w}_{n,\alpha}(x)w_{n,\alpha}(x^\prime)=\delta(x-x^\prime)$ and can be chosen real\cite{kohn},
one can look for a solution of nonlinear problem (\ref{case3nls}) in the form
$ \psi (t,x)=\sum_{n\alpha}c_{n,\alpha }(t)w_{n,\alpha}(x)$
where the expansion coefficients solve the set of equations
\begin{eqnarray}
i{\frac{dc_{n, \alpha}}{dt}} = \sum_{n_1}c_{n_1,
\alpha} \hat{\omega}_{n-n_1,\alpha } + 
 2\sigma \sum_{\alpha_1, \alpha_2, \alpha_3}\sum_{n_1,n_2,n_3}
  \bar c_{n_1,\alpha_1} c_{n_2,\alpha_2} c_{n_3,\alpha_3} W^{n n_1 n_2
n_3}_{\alpha \alpha_1 \alpha_2 \alpha_3 }
\label{exact}
\end{eqnarray}
with the nonlinear coefficients given by
\begin{equation}
W^{n n_1 n_2 n_3}_{\alpha \alpha_1 \alpha_2
\alpha_3} = \int_{-\infty}^{\infty}
w_{n,\alpha}w_{n_1,\alpha_1}w_{n_2,\alpha_2}w_{n_3,\alpha_3} dx.
\label{overlapp}
\end{equation}

In its general form, Eq. (\ref{exact}) is equivalent to (\ref{case3nls}) and thus it is not solvable. 
However, it allows reductions to  a simpler lattice in a number of important particular cases.
Indeed, $w_{n,\alpha}(x)$ can be chosen to be  localized around $x=n\tilde \Lambda$ and  exponentially decaying. 
If the OL depth is large enough, among all the coefficients $W^{n n_1 n_2 n_3}_{\alpha \alpha_1 \alpha_2 \alpha_3}$ those with $n=n_1=n_2=n_3$ are dominant and other terms can be neglected. Moreover, in this case the lowest zones are very narrow and the Fourier coefficients $\hat{\omega}_{n,\alpha}$ rapidly decay $|\hat{\omega}_{0,\alpha}|\gg|\hat{\omega}_{1,\alpha}|\gg
|\hat{\omega}_{n,\alpha}|$, ($n>1$).  Then restricting consideration to the only band $\alpha$ one reduces (\ref{exact}) to the discrete nonlinear Schr\"{o}dinger equation\cite{AKKS} (DNLS)
\begin{eqnarray}
i \frac{dC_{n}}{dt} =   \hat{\omega}\left(C_{n-1} +C_{n+1}\right) +\sigma |C_{n}|^2 C_{n}
\label{TB}
\end{eqnarray}
where $C_n= c_{n,\alpha}\exp(-i\hat{\omega}_{0,\alpha}t)\sqrt{2W^{nnnn}_{1111}}$ and $\hat{\omega}=\hat{\omega}_{1,\alpha}$. In the BEC theory DNLS (\ref{TB}) is referred to as the {\em tight-binding approximation}\cite{TrombSmerz,ABDKS,AKKS}. 

Now one can obtain the simplest stable solution which represents strong density localization near one of the potential minima, say near $n=0$. 
To this end we restrict the consideration to $\sigma>0$ and notice that considering the first zone $\alpha=1$ already for $V_0/E_{R}=-5$ we have\cite{AKKS} $\hat{\omega}_{0,1}\approx -2.1152$, $\hat{\omega}_{1,1}\approx -0.0192$ and $\hat{\omega}_{2,1}\approx 0.0002$, and thus  $\hat{\omega}_{n,1}$ can be regarded as a small parameter already for $n>1$. 
Next we fix the maximal density $|C_0|^2=1$ and look for a symmetric solution, $C_{-n}(t)\equiv C_n(t)$ in the form of the expansion: 
$C_n=e^{-i\Omega t}\sum_{j=n}^{\infty}C_n^{(j)}\hat{\omega}^j$ where $\Omega=\sum_{j=0}^{\infty}\Omega^{(j)}\hat{\omega}^j$. Direct substitution of this expansion into (\ref{TB}) allows us  to compute\cite{ABDKS}: $\Omega=1+2\hat{\omega}^2+O(\hat{\omega}^6)$ and $C_j=e^{-i\Omega t}\hat{\omega}^j+o(\hat{\omega}^{j+1})$.  Thus the decay of the atomic population is exponential: $|C_n|\propto e^{-\eta|n|}$, where $\eta=-\ln\hat{\omega}$ (a thin line in Fig.~\ref{fig:WF}b). For the above values of the relation between the potential energy amplitude and the recoil energy about $\approx 90\%$ of the condensed atoms are concentrated in the vicinity of the central site. Taking into account that the shape of a discrete soliton is well approximated by a WF, such discrete solitons can be called {\em Wannier solitons}\cite{AKKS}.

\subsection{Matter solitons. General approach}
\label{ms_ga}

The tight-binding approximation is valid when the lowest allowed zones are very narrow. Then one can approximate $\hat{\omega}_{0,1}\approx{\cal E}^{(1)}$ and $\hat{\omega}_{0,2}\approx{\cal E}^{(2)}$. 
Hence  shift of the energy of a discrete (Wannier) soliton  toward the forbidden zone is not small. 
Then a natural question about transformation between these gap solitons and discrete solitons, when the energy changes, arises. 

To answer this question we following Ref.\cite{AKS} 
look for a solution of (\ref{case3nls}) in the form $\psi(x,t)=e^{-i{\cal E} t}\varphi(x)$ which leads us to the equation 
\begin{eqnarray}
\label{differeq}
\varphi_{xx}+({\cal E} -U(x) - 2\sigma |\varphi|^2) \varphi=0
\end{eqnarray} 
where ${\cal E}$ plays now a role of a spectral parameter. 
Next, we separate the amplitude $u(x)$ and the phase $\theta(x)$ of $\varphi(x)$:  $\varphi(x)=u(x)e^{-i\theta(x)}$.
By substituting this expression into Eq.~(\ref{differeq}) we obtain the link $u^2\theta_x\equiv {\rm v}$ and the equation for $u(x)$: 
\begin{eqnarray}
u_{xx}-\frac{{\rm v}^2}{u^3}+({\cal E}-U(x)-2\sigma u^2)u=0. 
\label{RhoMain}
\end{eqnarray}
The constant ${\rm v}$ can be interpreted
as a normalized component of the condensate velocity, because ${\rm v}\propto (\bar{\psi}\psi_x- \psi\bar{\psi}_x)$. 

Due to the second term in Eq. (\ref{RhoMain}), $u(x)$
can vanish only for the zero velocity ${\rm v}=0$. Thus any stationary solution acquiring zero value at any point, including bright matter solitons (they have zero boundary conditions) can be excited only with a phase independent on $x$. Without restriction of generality in what follows we consider $\theta (x)\equiv 0$ and thus $\varphi(x)= u(x)$, where $u(x)$ solves the equation 
\begin{eqnarray}
u_{xx}+({\cal E}-U(x)-2\sigma u^2)u=0. 
\label{equ}
\end{eqnarray}

Eq. (\ref{equ}) can be interpreted as a Hamiltonian system with one and a half degrees of freedom. 
Let us, following\cite{AKS}, consider periodic potential (\ref{pot_cos}). Since Eq. (\ref{equ}) is invariant under shifts of the period of the potential, one can describe the solution by studying the area preserving Poincar\'e map ${\bf T}: (u _x(x),u (x))\longrightarrow (u _x(x+\pi),u (x+\pi))$ in the plane $(u _x,u )$.
The $n\pi$-periodic solutions of Eq.(\ref{equ}) correspond to the orbits of period $n$ of the map ${\bf T}$, i.e. to fixed points of the map ${\bf T}^n$. These fixed points can be either hyperbolic or elliptic\cite{GH}.
Solitons correspond to homoclinic and heteroclinic orbits of the Poincar\'e section, i.e. to sequences of points $\{R_n=(u ^\prime_n,u _n),n\in\mathbb{Z},u_n=u(x_0+\pi n), u_n^\prime =\left.u_x
\right|_{x=x_0+\pi n}\}$, $R_{n+1}={\bf T}R_n$, such that $R_n$ tends to the same (homoclinic orbit) or different (heteroclinic orbit) fixed points of $T$ as $n=\pm \infty$. 
Each hyperbolic fixed point ${\cal O}$ has stable, $W_s({\cal O})$, and unstable, $W_u({\cal O})$, manifolds, whose intersection means the existence of a homoclinic orbit.
If the map ${\bf T}$ is defined in the entire plane and is smooth, a single transversal intersection of these manifolds implies the existence of infinitely many homoclinic orbits of ${\bf T}$ that may be coded by Bernoulli sequences\cite{GH}. Properties of ${\bf T}$ depend on the sign of $\sigma$. 
For $\sigma<0 $, the map is well defined in the whole plane, so that one expects the existence of
{\em infinitely many} stationary soliton-like solutions of Eq.(\ref{equ}). 
On the contrary, for $\sigma>0$, the solutions of Eq. (\ref{equ}) may blow up in modulus due to the sign and the power of the nonlinearity, so that the map ${\bf T}$ cannot be defined in the entire plane $(u_x,u )$. 
In this case a number of solitons is drastically reduced. 

Bright matter solitons correspond to solutions of Eq.(\ref{equ}) that vanish as $x\to\pm\infty$. 
In terms of the dynamical system generated by the map ${\bf T}$, they are linked to
homoclinic orbits of the fixed point ${\cal O}=(0,0)$, which  must be hyperbolic. This happens when
${\cal E}$ lies in a forbidden zone of (\ref{Bloch}). Here we restrict consideration to the first forbidden zone:
${\cal E}^{(1)}<{\cal E}<{\cal E}^{(2)}$. Eq.(\ref{Bloch}) has only one
solution (up to constant factors) which decays as $x\to\infty$:
$\varphi_0(x)=P(x)e^{\mu x}$, where $P(x)=P(x+2\pi)$ and $\mu$ the Floquet exponent. In this
case, the eigenvalues of the map ${\bf T}$ at the point $(0,0)$ are $\lambda_{1,2}=e^{\pm\mu\pi}$. From Eq. (\ref{Bloch}) it follows
that the asymptotic behavior of the bright soliton solution is
given by the expression $u (x)\sim G \varphi_0(x)$, where $G$ is a real constant, what allows numerical constructing the solutions of (\ref{equ}). 
 It was found in Ref.\cite{AKS} that there exist two (up to sign) simple states (even and odd) centered in a maximum (minimum) or in a zero of the lattice potential. All other bright soliton solutions can be interpreted as bound states of these elementary entities. 
 
\begin{figure}[h]
\vspace*{8pt}
\centerline{\epsfig{file=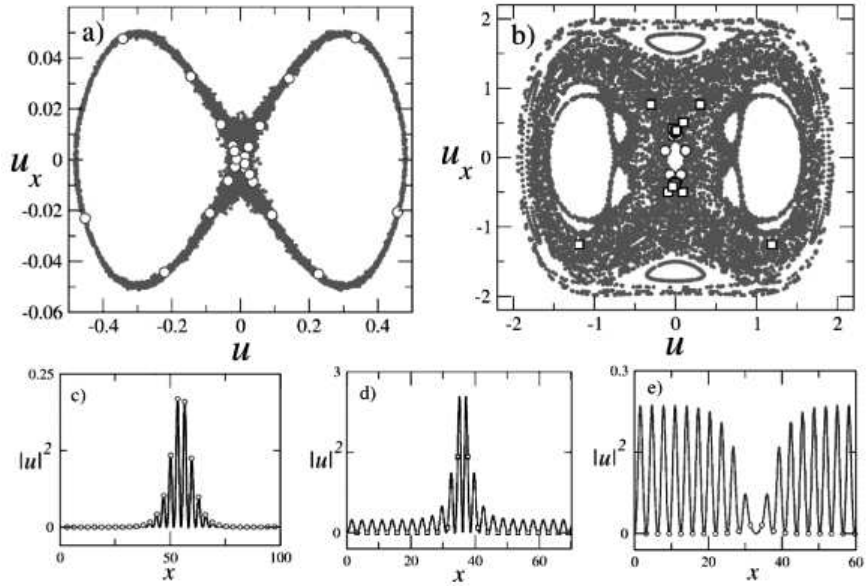,width=12cm}}
\vspace*{8pt}
\caption{Phase portraits obtained by iterating  
${\bf T}$ for Eq. (\ref{equ}) with the parameters: (a)  $A=-1$, ${\cal E}=0.55$, $\sigma=1/2$ and (b)  $A=1$, ${\cal E}=0.3$, $\sigma=-1/2$.
(c) An example of a bright soliton corresponding to open circles in (a); Examples of a bright (d) and  dark (e) solitons against a periodic background (squares and circles in (b) correspondingly).
} \label{figtwo}
\end{figure}

Fig.~\ref{figtwo}a represents a stochastic layer obtained by
iterating the Poincar\'e map ${\bf T}$ starting with an initial point close to $(0,0)$ (about half a million points). The opened circles on top of the stochastic layer are the points $\{u^\prime_n,u_n\}$  that correspond to the bright soliton solution of Eq. (\ref{equ}), explicitly shown in Fig.~\ref{figtwo}c. When the energy ${\cal E}$ is lowered to the bottom of the gap, ${\cal E}^{(1)}$, the stochastic layer becomes thin and the amplitude of the corresponding solution tends to zero, what corresponds to the solution (\ref{bright}) obtained in the weakly nonlinear limit (Section~\ref{meanfield}). When the energy approaches the upper boundary of the gap, ${\cal E}^{(2)}$, the soliton amplitude growth. Since the gap width is finite there exist upper bounds for the   amplitude  of a matter soliton and for a number of particles constituting it.

Matter solitons on a periodic background also  can be constructed using the dynamical system generated by the
map ${\bf T}$. They correspond to homo- and heteroclinic orbits of hyperbolic fixed points of the maps ${\bf T}^n$ ($n=1, 2, \ldots$), other than the point $(0,0)$.
Consequently, the background  must be only  a 
{\em hyperbolic}  periodic solution of  Eq. (\ref{equ}).
It has been found in Ref.\cite{AKS} that there exists
a set of $\pi$- and $2\pi$-periodic solutions of Eq. (\ref{equ}). 
In the case  $\theta(x)\equiv 0$, one can find  solutions that are characterized by Fourier expansions with either cosine or sine terms (up to $\pi/2$-shifts). 

Each periodic solution of the hyperbolic type can be related to {\em four} simplest soliton solutions against its background. Two of them are bright and dark symmetric solutions, i.e. having a maximum or a minimum of amplitude at the center (the former are also referred to as anti-dark solitons), and the two others are bright and dark antisymmetric solitons. 
In the case $\sigma<0$, we have observed that all the four soliton solutions may {\em coexist} for some range of the values of ${\cal E}$, $A$, and the amplitude of the background periodic solution. They may also bind together and form complex multisoliton states. 
The situation is different for $\sigma>0$. 
In this case, there exist {\em two kink-shaped solitons}, which can be treated as transitions between two periodic backgrounds of different sign.
An example of numerical study of the Poincar\'e map ${\bf T}$ corresponding to existence of a dark and anti-dark solitons is shown in Fig.~\ref{figtwo}b,d, and e.
These solutions are counterpart of bright and dark solitons against a background in the theory of DNLS equation\cite{CKV} (\ref{TB}).
  
In the present text we do not discuss stability of matter solutions. In the case of small-amplitude solitons the evolution equation is obtained by excluding secular terms and thus guarantees stability in some sense. In a more general case the stability can be established by the direct numerical simulations or by providing the linear stability analysis\cite{LOSK,KCGTFM}. We also notice that the dynamical system approach can be applied directly to a lattice approximation to the GP equation\cite{CGP}. 
Meantime, a complete classification of all available matter soliton solutions like this was recently done\cite{ABK} for the discrete intrinsic localized modes of the DNLS model  remains to be an opened problem.
\section{Nonlinear periodic waves}
\label{periodic}

As it is well known,\cite{IST} NLS equation (\ref{nls1D}) possesses exact periodic wave solutions which, also like solitons, can be obtained by means of the inverse scattering method. 
In our case this means that one can excite not only localized matter waves but also nonlinear periodic matter waves. 
Periodic waves survive in the limit when the healing length becomes of order of the lattice constant, i.e. in the case of the evolution equation (\ref{nls1dcase2}), what has been shown in Section~\ref{soliton}. A remarkable fact is that for some particular potentials one can find exact analytical periodic solutions valid for the whole range of the relation between the lattice constant and the healing length. This is the case of potential (\ref{eq:pot}), studied in detail in\cite{24} and briefly summarized in the present section. 

Let us focus on stationary solutions of Eq. (\ref{case3nls}). In contrast to bright matter solitons, periodic waves may do not acquire zero values on the real axis, and then they may posses $x$-dependent phases (the so-called nontrivial phase solutions). If the phase is constant (then it can be put  zero) the solution is referred to as a {\em trivial phase solution}, and it solves Eq. (\ref{equ}). Here we discuss only the last case.

We observe that if stationary periodic solutions of the NLS equation (\ref{nls1D}) with $M =1/2$ are known, we denote them $\psi=e^{-i{\cal E}t}u_0(x)$, one can easily verify, that the function $\psi=e^{-i{\cal E}t}\tilde{u}(x)$ with $\tilde{u}(x)=\sqrt{\chi}u_0(x)$, $\chi>0$, solves (\ref{case3nls}) with a periodic potential $U(x)=2\sigma (1-\chi)|u_0(x)|^2$. Alternatively,  Eq. (\ref{equ}) allows one to construct a lattice potential $U(x)$, in an explicit form, when the desired wave field $u_0(x)$ is given, subject to the requirement  of a regular behavior of $u_{0,xx}/u_0$: $U(x)=\frac{u_{0,xx}}{u_0}+{\cal E}-2\sigma u_0^2$. In particular, one can verify that this is the case of the Jacobi elliptic functions pq$(\kappa x,m)$, where p=c,s or d, and ${\rm q=n}$ (more generally one can consider, say, 	$u_{\rm pq}(x)=\left(A\, {\rm pq}^\beta(\kappa x,m)+B\right)^\alpha$ with $B> |A|$, $\alpha$ and $\beta$ being real integers). Respective solutions, which subject to proper choice of ${\cal E}$ lead to potential (\ref{eq:pot}), are as follows\cite{24}:
\begin{eqnarray}
	\label{sn_a}
	\psi_{\sn}(x,t)&=&e^{i\kappa^2(1+m)t}u_{\sn}(x),\quad 
	u_\sn=\sqrt{\left(\kappa^2m-\frac{V_0}{2}\right)
	\sigma}\, \sn(\kappa x\mid m),
	\\
	\label{cn_a}
	\psi_{\cn}(x,t)&=&e^{i[\kappa^2(2m-1)-V_0]t}u_\cn(x),\,\,\,
	u_\cn(x)=\sqrt{\left(\frac{V_0}{2}-\kappa^2m\right)\sigma}
	\, \cn(\kappa x\mid m),
	\\
	\label{dn_a}
	\psi_{\dn}(x,t)&=&
	e^{i[\kappa^2(2-m)-\frac{V_0}{m}]t}u_\dn(x),\,\,\,
	u_\dn(x)=\sqrt{\left(\frac{V_0}{2m}-\kappa^2\right)\sigma}	 
	\,\dn(\kappa x\mid m),
\end{eqnarray}
where the parameters $\kappa$, $\kappa>0$, and $m$, $0\leq m\leq 1$, must be chosen to make an expression under the square root positive.
 
In the absence of the lattice potential, $V_0=0$, the sn-wave exists in a BEC with a positive scattering length, $\sigma=1$, while  cn- and dn-waves exist in BECs with negative scattering length, $\sigma=-1$. They are  solutions of the NLS equation (\ref{nls1D}) with $M=1/2$. The trap potential changes the situation, allowing existence of sn-waves (cn- and dn-waves) in BECs with negative (positive) scattering lengths.
 
If one considers the limit $m\to 1$ at $V_0=0$ (or other allowed values of the potential amplitude), one obtains the transitions from the periodic waves to the solitons, given by (\ref{bright}) and (\ref{dark}) with $M=1/2$ and $\kappa=\eta$: $\psi_{\sn}\to\psi_{{\rm ds}}$, $\psi_{\cn}\to\psi_{{\rm bs}}$, and $\psi_{\dn}\to\psi_{{\rm bs}}$. This means that by manipulating the parameter $m$ one can  provide transformation among different solutions (see Section~\ref{FR}).

Finally it worth to mention that the linear case,   corresponding to very low densities, i.e. to the limit of zero amplitudes, is achieved when $m\to \frac{V_0}{2\kappa^2}$, what is possible when $V_0\geq 0$. In this case, the above solutions are transformed into the particular Bloch functions of the potentials $V_L=V_0\sn^2\left(\kappa x\mid\frac{V_0}{2\kappa^2}\right)$.

\section{Management of matter waves by Feshbach resonance}
\label{FR}

When a periodic wave is given by one of the Jacobi elliptic functions one can provide deformation of this wave to a sequence of localized matter waves by modulating parameters $m$ and $\kappa$. On the other hand, in experimental\cite{bright,FR} and theoretical\cite{FR_theory,Moer} works it was shown that an effective tool for manipulating BECs is the {\em Feshbach resonance} (FR), allowing one to change the scattering length by varying external magnetic field $B(t)$ near the resonant point $B_0$ (Ref.\cite{Moer}):
$ a_{s}(t) = a_s(0) \left(1 + \frac{\Delta}{B_{0}- B(t)}\right)$ (here $a_s(0)$ is the asymptotic value of the scattering length far from resonance, and  $\Delta$ is the width of the resonance).
Since change of the  nonlinearity results in change of $m$ (see below),  trains of solitons can be generated from a periodic wave in a controllable manner\cite{prl90AKKB}.

Let us consider the problem of soliton generation in more details (generalizing earlier results\cite{prl90AKKB} to the case of a BEC in an OL). Temporal dependence of $a_s$ means the dependence on time of the nonlinear coefficient: $g_0\to \tilde{g}_0(\epsilon^2t)\equiv g_0\cdot g(t_2)$ where we  assume slow dependence on time and normalization $g(0)=1$. Then one can repeat the arguments of Section~\ref{meanfield}, substituting $g_0$ by $\tilde{g}_0(t_2)$. The modification of each of the evolution equations   (\ref{nls1D})--(\ref{nls1_case4}) is reduced to the modulation of the nonlinearity by the function  $\tilde{g}(t)$. In particular, (\ref{case3nls}) is reduced to
\begin{equation}
\label{NLS-F}
i\psi_t+ \psi_{xx}-  U(x)\psi-  2\sigma g(t)|\psi|^2\psi=0.
\end{equation} 

If $g(t)$ is a positive function (the only case considered below) by the ansatz
\begin{equation}
\label{u-v}
  \psi(x,t)=\frac{v(x,\zeta(t))}{\sqrt{g(t)}}= e^{-\zeta(t)/2} v(x,\zeta), 
\end{equation}
where $  \zeta\equiv \zeta (t)=2\int_0^t\gamma(t') dt'=\ln g(t)$, $ \zeta(0)=0 $, and $v(x,0)=\psi(x,0)$
Eq. (\ref{NLS-F}) is transformed into a dissipative NLS equation 
\begin{equation}
\label{modNLS}
  iv_t+v_{xx}- U(x)v-2\sigma |v|^2v=i\gamma (t) v
\end{equation}
where $\gamma(t)=g_t/2g =\zeta_t/2$.
Thus, for slowly varying $g(t)$, $|\gamma(t)|\ll 1$ and the right hand side of Eq.~(\ref{modNLS})
can be considered as a small perturbation. 
 
Let us focus on perturbed evolution of periodic waves assuming that initially they are given by $v(x,0)=u_{{\rm pq}}(x)$ where $u_{{\rm pq}}(x)$ is one of the functions given by (\ref{sn_a}), (\ref{cn_a}) or (\ref{dn_a}). Under influence of the dissipative perturbation  a wave shape is changed what can be described by variations of the parameters $m$ and $\kappa$  assuming them to be slow functions of time $t$ (the {\em adiabatic approximation}).  To this end we observe that  the coefficients in the dissipative model
(\ref{modNLS}) are independent of $x$, and thus the perturbation does not result in change of the period $\Lambda$ of the nonlinear wave what can be expressed as follows
\begin{equation}
\label{wl}
  \frac{d}{d\zeta}\Lambda(m,\kappa)=0.
\end{equation}
Next it is a straightforward algebra to ensure that the integral 
$N=\int_{0}^{\Lambda} |v|^2dx$ is governed by the equation
\begin{equation}\label{Neq}
 \frac{dN(m,\kappa)}{d\zeta}= N(m,\kappa).
\end{equation}

If the expressions for $\Lambda$ and $N$ in terms of $m$ and $\kappa$ are known, Eqs.~(\ref{wl}) and (\ref{Neq}) reduce to a system of two differential equations of the first order for $m$ and $\kappa$. 
The form of this system depends, of course, on the choice of the initial ${\rm pq}$-wave. 
Below, we concentrate on two examples of transformations of  periodic waves in trains of bright and dark solitons (for more details see Refs.\cite{VVKBook,prl90AKKB}).

\subsection{${\rm dn}$-wave in a BEC with a negative scattering length.}

Consider a BEC with a negative scattering length, $\sigma=-1$, which has initial form of the dn-wave $u_{{\rm dn}}(x)$, given by Eq. (\ref{dn_a}). We compute
$	\Lambda=2{\rm K}(m)/\kappa,$ and $N=(2\kappa^2m-V_0){\rm E}(m)/(\kappa m)$, and from Eqs. (\ref{wl}), (\ref{Neq}): 
\begin{eqnarray}
\label{dn-adiab-eqn}
\begin{array}{l}\displaystyle{
\frac{d\kappa}{d\zeta}=\frac{\kappa ( 2\kappa^2m-V_0)}{\Delta}{\rm E}(m)\left[{\rm E}(m) - m^\prime{\rm K}(m)\right],}\\
\displaystyle{
\frac{dm}{d\zeta}=\frac{2(2\kappa^2m-V_0)mm^\prime}{\Delta}{\rm E}(m){\rm K}(m)}
\end{array}
\end{eqnarray}
where $\Delta=V_0\left[{\rm E}^2(m) + m^\prime{\rm K}^2(m)\right]+2k^2m\left[{\rm E}^2(m)-m^\prime{\rm K}^2(m)\right]$ and $m^\prime=1-m$ is a standard notation. 

\begin{figure}[ht]
\vspace*{8pt}
\centerline{\epsfig{file=dnv01.eps,width=4cm}
\epsfig{file=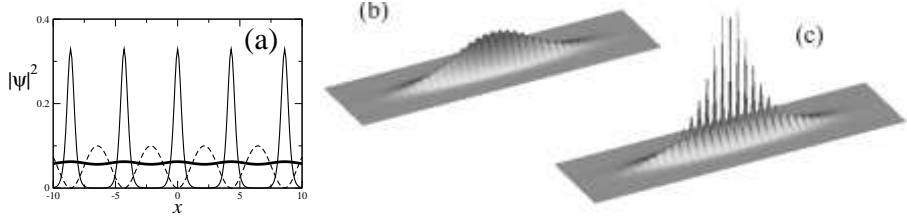,width=8cm}}
\vspace*{8pt}
\caption{(a) Adiabatic modulation of a dn-wave solution with $m(0)=0.11$ and $\kappa(0)=0.75$ at
 $\zeta=0$ (thick solid lines) and at $\zeta=3$ (thin solid  lines) in potential (\ref{eq:pot}) (dashed lines) with $V_0=0.1$. 
(b) Initial ($t=0$) and (c) final ($t=6$~ms) density profiles obtained by numerical solution of (\ref{3Dgp}) with   $\nu=0.02$ and $N\approx 10^4$ atoms of  $^{7}$Li.    
Initially the dn-wave has the same parameters as in (a) but now modulated by the Gaussian function $\exp(\nu x^2/2+r^2/2)$.  The amplitude of OL is $V_0=0.2 E_R$. The size of the image is $200\times 80\, \mu$m.  The characteristic times of FR are $\tau=2$~ms and $T_0=3$~ms (see the text).  
}
\label{dn-pot}        
\end{figure}

In Fig.~\ref{dn-pot}a we present a numerical solution of the system (\ref{dn-adiab-eqn}) which illustrates creation of a train of bright solitons in the minima of the OL. 

To check the adiabatic approximation we solved numerically (Fig.\ref{dn-pot} b, c) the dimensionless 3D GP equation for cigar shaped condensate with cylindrical symmetry and $\sigma=1$: 
\begin{eqnarray}
\label{3Dgp}
i\partial_t \psi=-\left(\partial^2_x +\frac{1}{r}\partial_rr\partial_r\right)\psi
+ V_{ext}(x,r)\psi+2 e^{t/\tau} |\psi|^2\psi\, \label{3DGP_dim}
\end{eqnarray}
where $r^2=y^2+z^2$ and  $V_{ext}(x,r)=r^2+\nu^2 x^2+ V_0 \kappa^2 \sn^2(\kappa x\mid m)$.
We switch off the FR at $T_0$ and in the further calculation we take $g=g(T_0)$.   
One again can clearly see that after some time a set of bright pulses localized in the minima of the OL potential arises. These pulses represent solitons, which would survive for a sufficiently long time after the lattice potential is switched off.

Similar results of generation of trains of bright solitons in a BEC with repulsive interactions can be obtain by considering cn-wave (\ref{cn_a}) as an initial profile of the condensate.


\subsection{Evolution of a ${\rm sn}$-wave in a BEC with a positive scattering length.}

As a second example we consider deformation of the sn-wave (\ref{sn_a}) in a BEC with a positive scattering length, $\sigma=1$ affected by the FR, which leads to creation of a train of dark solitons. Following the steps of the adiabatic approximation now one can compute $\Lambda=4{\rm K}(m)/\kappa$, $N=2(2\kappa^2m-V_0)\left[{\rm K}(m)-{\rm E}(m)\right]/(\kappa m)$ and deduce from (\ref{wl}), (\ref{Neq}): 
\begin{eqnarray} \label{sn-pot-ad1} 
 \begin{array}{l}\displaystyle{
\frac{d\kappa}{d\zeta}=\frac{\kappa(2\kappa^2m-V_0)}{\Delta}
\left[{\rm E}(m)-{\rm K}(m)\right]\left[{\rm E}(m)-m^\prime{\rm K}(m)\right],}\cr
	\displaystyle{
\frac{dm}{d\zeta}=\frac{2mm^\prime(2\kappa^2m-V_0)}{\Delta}{\rm K}(m)\left[{\rm E}(m)-{\rm K}(m)\right]}.
 \end{array}
\end{eqnarray}
Here $\Delta=V_0\left[{\rm E}^2(m)-m^\prime{\rm K}^2(m)\right] 
+2\kappa^2m\left\{\left[{\rm E}(m)-{\rm K}(m)\right]^2 -m {\rm K}^2(m) \right\}$.


\begin{figure}[ht]
\centerline{\epsfig{file=sn.eps,width=5cm}\quad
\epsfig{file=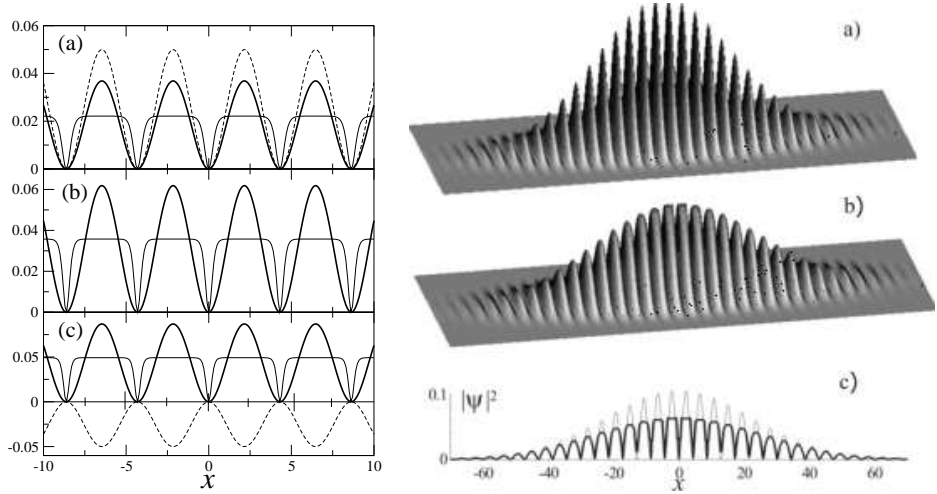,width=7cm}}
\vspace*{8pt}
\caption{Left panels: Adiabatic dynamics of sn-wave solutions at $\zeta=0$ (solid thick lines) and $\zeta=\zeta_{fin}$ (solid thin lines) in the periodic potential (dashed lines) for (a) $V_0=0.05$, and $\zeta_{fin}=5$; (b)  $V_0=0$, $\zeta_{fin}=6$, and (c)  $V_0=-0.05$, $\zeta_{fin}=6$.
Right panels: (a) Initial ($t=0$) and (b) final ($t=0.8$ ms) density profiles of the condensate embedded in 1D OL (\ref{eq:pot}) with strength $V_0\approx -0.2 E_R$, aspect ratio is $\nu=0.05$ and characteristic time of FR is $\tau\approx 0.06$~ms and $T_0\approx 0.45$~ms. Initial profile contains $N\approx 5\times 10^4$ $^{87}$Rb atoms.
The size of the images is $150\times 40\, \mu$m. (c) Crossection of the density $|\psi(r=0,x)|^2$ at initial (thin line) and final (thick line) times corresponding (a) and (b). In all cases $m(0)=0.11$ and $\kappa(0)=0.75$. }
\label{sn-pot}        
\end{figure}

The left panels of Fig.~\ref{sn-pot} show BEC dynamics in the adiabatic approximation, where emergence of dark holes, which can be associated with dark solitons can be observed. Notice that in the cases (a) and (c) dark solitons are created in the minima and maxima of the lattice potential, respectively.
A train of darks soliton can also be obtained in the 3D simulations of the model (\ref{3Dgp}) starting from the sn-wave as initial profile modulated by Gaussian function $\exp(\nu x^2/2+r^2/2)$. Like in the previous case varying nonlinearity results in creation of solitary pulses. However, in contrast to the case of bright solitons, if the parabolic potential is switched off, the condensate constituting the background for dark solitons, start to expand rapidly passing to the linear regime. 
\section{Bloch oscillations}
\label{bloch}

In the preceding sections it was shown that OLs can be employed to create solitary and periodic matter waves in a controllable manner. In this and next sections we concentrate on the use of OLs for manipulating matter solitons. We start with the phenomenon very well known in the solid sate physics\cite{solid} -- Bloch oscillations  -- consisting in periodic motion (in direct and reciprocal spaces) of a quantum particle in a periodic potential. Since an  atom in an OL undergoes Bloch oscillations\cite{Dahan}, BEC must also undergo Bloch oscillations when loaded in an OL, what has indeed been observed experimentally\cite{Ander,LZ,MorIng} and studied numerically\cite{CN_99,Bragg}. In these studies a BEC was considered as an atomic cloud, where each atom undergoes Bloch oscillations. However, {\it a priori} it is not evident whether this phenomenon can be observed with matter solitons (considered as single entities). Indeed, as it follows from the effective mass approach (Section~\ref{meanfield}), when in reciprocal space the condensate passes from one BZ edge to another, the effective mass changes its sign, and thus a bright matter soliton existing at one edge becomes a dispersive wave packet at another edge\cite{Bragg,dispersion}.  

It turns out, that Bloch oscillations, although in a somehow modified sense, can indeed be observed\cite{ABDKS}.  Mathematically this possibility is explained by two facts: lattice solitons subject to a linear force undergo Bloch oscillations\cite{Bloch,SB} and strongly localized matter solitons in the tight-binding approximation are well described by the DNLS equation (see Section~\ref{soliton}). 

\begin{figure}[h]
\centerline{\psfig{file=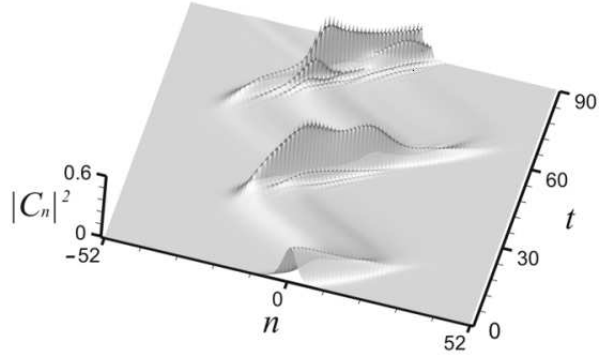,width=8cm}}
\vspace*{8pt}
\caption{Dynamics of the solitary wave $C_n(t)=\sqrt{2}\sinh(\beta) \mbox{sech}(\beta
n)\exp(-i\alpha n)$  affected by the linear force
$\hat g=16/l_{chain}$ with $l_{chain}=104$, $\hat\omega=1$ and $\sigma=1$. Initial soliton parameters $\beta=0.4$,
$\alpha=1.571$.}
\label{fig_bloch}
\end{figure}

To illustrate the phenomenon we assume that a BEC is loaded in a vertical OL, like in the experimental setting\cite{Ander}, and thus besides the lattice potential, a gravitational field, is applied to the condensate. Alternatively, one can consider an accelerating lattice, where the linear $x$-dependent force appears due to the acceleration\cite{CN_99}. The gravitational force results in the term $mgx\Psi$ (notice that now $x$ is a vertical axis, and $g$ is the gravitational acceleration) in the  GP equation (\ref{GPE}). If one also considers relatively large amplitude of the lattice potential preventing Landau-Zener tunneling between bands\cite{LZ}, then in the tight-binding approximation (\ref{TB}) a new term  $\hat{g}n c_{n,\alpha}$, where $\hat{g}n=\frac{ma_\bot}{\hbar}{\omega_\bot}g$, is to be added:    
\begin{eqnarray}
	\label{TB-B}
	i \frac{dC_{n}}{dt}  	= \hat{\omega}\left(C_{n+1} +C_{n-1}\right)+\hat{g}nC_n+\sigma |C_{n}|^2 C_{n}.
\end{eqnarray}
The obtained system was studied numerically in Ref.\cite{SB}, where it has been discovered that the bright solitary pulse   undergoes periodic motion, shown in Fig~\ref{fig_bloch}. 

Analytically, the effect can be understood in approximation as follows. Introduce momentum $P=\sum_np_n$ and energy $E=\sum_ne_n$ of the soliton, where $p_n=(C_n\overline{C}_{n-1}-C_{n-1}\overline{C}_{n})/(2i)$ and $e_n=(C_n\overline{C}_{n-1}+C_{n-1}\overline{C}_{n})/2$. It is a straightforward algebra to ensure that
\begin{eqnarray}
\label{PE}
  \begin{array}{l}
	\displaystyle{
	\frac{dP}{dt}=
	\hat{g}E+\sigma\sum_n(|C_n|^2-|C_{n-1}|^2)e_n,
	}
	\\
	\displaystyle{ 
	\frac{dE}{dt}=
	-\hat{g}P-\sigma\sum_n(|C_n|^2-|C_{n-1}|^2)p_n.
	}
	\end{array}
\end{eqnarray}
Thus if populations of the neighbor sites of the OL, defined by $|C_n|^2$, are not very dispersed $\mid |C_n|^2-|C_{n-1}|^2\mid \ll \hat{g}$, one can neglect the last terms in the both equations (\ref{PE}), and obtain that $P$ and $E$ periodically vary with the frequency $\hat{g}$ and with the ``dispersion relation'' $P^2+E^2=P_0^2+E_0^2$ where $P_0$ and $E_0$ are the initial values of the momentum and the energy of the soliton.

\section{Matter solitons in inhomogeneous lattices}
\label{defect}

So far we dealt with homogeneous lattices. It turns out that by modulating an OL one can efficiently manage matter solitons as they accelerate, decelerate, oscillate or undergo the reflection depending on the type of the modulation introduced\cite{BKK}.  In particular lattice modulations can simulate attractive and repulsive defects in the gap soliton dynamics. 

Let us consider two examples, of a BEC with a negative scattering length, $\sigma=-1$, where the periodic potential is given by one of the expressions as follows
\begin{eqnarray}
\label{def1}
	V_L({x})=\left[1\pm e^{-\epsilon^{5/2} ({x}-\Delta {x})^2}\right] 
\cos(2{x}). 
\end{eqnarray}
 
\begin{figure}[h]
\centerline{\epsfig{file=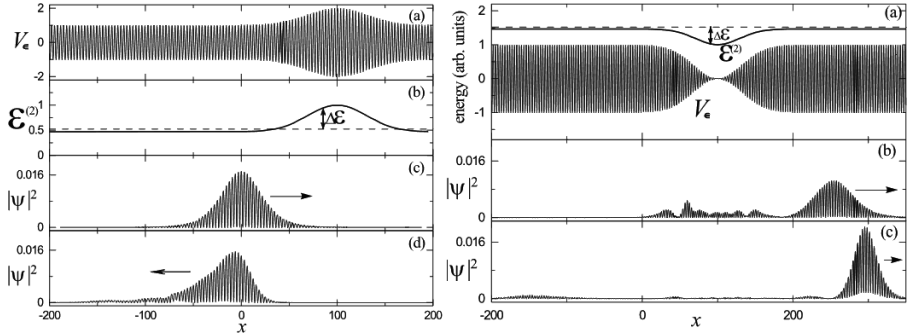,width=12cm}}
\vspace*{8pt}
\caption{Left panels: (a) Periodic potential with defect given by Eq.~(\ref{def1}) with upper sign and parameters $\epsilon=0.05$ and $\Delta {x}=100$. (b) The upper edge of the lowest gap (solid line) and the energy of the soliton with the initial velocity $v=0.1$ (dashed line). Intersections of these lines represent the reflection points.  (c) Initial profile and (d) profile of the reflected soliton   at 
time $t=150$.
Right panels: (a) Periodic potential with the defect given by Eq.~(\ref{def1}) with lower sign and corresponding upper edge ${\cal E}^{(2)}$ of the first band gap (here the dashed line is initial soliton energy). Parameters of the defect are the same as in left panel. Profiles of the soliton with different initial velocities $v=0.1$ (b), and  $v=0.3$ (c) at times $t=300$ and $t=190$,  respectively.
}
\label{fig5}
\end{figure}

Fig.~\ref{fig5} shows the both types of the lattice modulations ((a) in left and right panels) and the respective soliton dynamics (left panel (c), (d) and right panel (b) and (c)).
In both cases we start with numerically obtained small-amplitude bright solitons which exist in the vicinity of the upper band gap edge ${\cal E}^{(2)}$ with ${\cal E}^{(2)}-{\cal E}<0$  (see Fig.\ref{modinstab} and also Section~\ref{ms_ga}) ensuring that the soliton has a small initial velocity.  
  
The left set of panels of Fig.~\ref{fig5} shows reflection of a matter soliton by the defect.  This phenomenon can be interpreted as a Bragg reflection as it is explained in Fig.~\ref{fig5}b. Indeed, smooth modulation of the lattice results in a deformation of the gap edges in such manner that the upper gap edge has bigger energy (${\cal E}^{(2)}+\Delta{\cal E}$) than the energy of the gap edge of the unperturbed lattice, 
${\cal E}^{(2)}$ (see Fig.~\ref{fig5}b). Thus a soliton moving from the left to the right faces the boundary of the gap where its velocity becomes zero. The soliton is reflected from the defect, which can be viewed as a repulsive impurity.

In the case of a local decrease of the potential depth, the forbidden gap is locally shrinks, the situation is described by the lower sign in (\ref{def1}) and illustrated in the right panels of Fig.~\ref{fig5}. In this case, in the region of the defect the soliton energy is effectively shifted from the gap edge ${\cal E}^{(2)}$ toward the allowed band (see Fig.~\ref{fig5}a), where the group velocity of the carrier wave has local maximum.  
  Thus the modulation described by the Eq. (\ref{def1}) with lower sign acts as an attractive defect, and some 
number of atoms could be captured by such a defect, what happens when an  initial kinetic energy of the 
condensate is small enough. This is exactly what we observe  in numerical simulations shown in 
Fig.~\ref{fig5}b (right panel). The higher velocity matter waves pass through the defect without substantial 
changes (see Fig.~\ref{fig5}c).
 
To conclude this section we notice that the above examples (for more details and other examples see Ref.\cite{BKK}) show that matter solitons in OLs can be manipulated by affecting the carrier wave background, something not possible to do with matter solitons in homogeneous BECs.
\section{Conclusion}
\label{conclusion}

In this brief review we have shown a diversity of dynamical effects which can be observed in BECs loaded in far-off-resonant OLs, the latter representing an efficient tool for managing matter waves. In particular, OLs allow one to obtain solitary and periodic matter waves, which cannot exist in homogeneous BECs, to deform solutions, to arrange their motion in an {\it a priori} given manner, and to manipulate matter soliton dynamics by affecting the dynamics of the carrier matter waves. 

Limited volume of the present work however, did not allow us to cover a number of other important issues related to BECs in OLs, even limiting consideration to the mean-filed approximation. Among them we mention matter waves in multicomponent\cite{multicom} and in atomic-molecular\cite{AMBEC} condensates, the effect of Landau-Zener tunneling\cite{LZ,KKS}, superfluidity\cite{superfluid},  delocalization of matter waves\cite{deloc}, dispersion management of matter solitons in OLs\cite{dispersion,AbdMal}, generation of matter shock-waves in OLs\cite{ABDKS}, propagation matter waves  through finite lattices\cite{CELR}, etc.  Finally, we have not discussed the relevant stability properties of periodic and localized matter waves.
\section*{Acknowledgement}

This work includes the results obtained by the present authors in fruitful collaboration with M. Salerno, G.L. Alfimov, F.Kh. Abdullaev, P.G. Kevrekidis, B.B. Baizakov, A.M. Kamchatnov, and V. Kuzmiak, who are gratefully acknowledged. For obtaining some of the numerical results presented here we used  numerical tools created by G.L. Alfimov.
 
Work of V.A.B. was supported by the FCT fellowship SFRH/BPD/5632/2001. 

\end{document}